\documentclass[aip,groupedaddress,reprint]{revtex4-1}

\usepackage{graphicx}
\usepackage{dcolumn}
\usepackage{bm}

\usepackage[utf8]{inputenc}
\usepackage[T1]{fontenc}
\usepackage{mathptmx}
\usepackage{amsmath}
\usepackage{amssymb}
\usepackage{xcolor}
\usepackage{soul}
\usepackage{gensymb}
\usepackage{lineno}

\begin{document}

\preprint{AIP/123-QED}

\title[Turing patterns vs chimera states in the Brusselator model]{From Turing patterns to chimera states in the 2D Brusselator model}

\author{A. Provata}
 \email{a.provata@inn.demokritos.gr}
\affiliation{ 
Institute of Nanoscience and Nanotechnology, National Center for Scientific Research ``Demokritos'', 15341 Athens, Greece }%

\date{\today}

\begin{abstract}
The Brusselator has been used as a prototype model for autocatalytic reactions, and in particular for the 
Belouzov-Zhabotinsky reaction. When coupled at the diffusive limit, the Brusselator undergoes a
Turing bifurcation resulting in the formation of classical Turing patterns, 
such as spots, stripes and spirals in 2 spatial dimensions. 
In the present study we use generic nonlocally coupled Brusselators
 and show that in the limit of the coupling range $R \to 1$ (diffusive limit),
the classical Turing patterns are recovered, while for  intermediate
coupling ranges and appropriate parameter values
chimera states are produced. This study demonstrates how the parameters
of a typical nonlinear oscillator can be tuned so that the coupled system passes 
from spatially stable Turing structures to 
dynamical spatiotemporal chimera states.
\end{abstract}

\maketitle
\everymath{\displaystyle}
\begin{quotation}
Oscillating reactions, also known as chemical clocks, have been studied intensively in the
past, with notable examples
the Bray-Liebhafsky reaction (1921), the Belousov-Zhabotinsky reaction (1958),
the Briggs-Rauscher reaction (1972) and the Chlorite-Iodide-Malonic Acid (CIMA) reaction (1990). 
They are characterized by periodical color changing
or periodic gas emissions and this periodic activity
 was earlier considered as a contradiction to common sense. 
The explanation of the mechanisms behind the unexpected 
concentration oscillations was proposed in the 1970's,
and relies on the nonlinear nature of the interactions between the species,
which give rise to a limit cycle following a Hopf bifurcation scenario \cite{epstein:1998}.
The same nonlinear mechanisms are responsible for the emergence of spectacular Turing
patterns when spatial diffusion is taken into account, as demonstrated by Alan Turing in his 1952 
seminal paper \cite{turing:1952}.  In the present study we quest the existence of a
different unexpected spatiotemporal effect, the chimera states, which are mainly observed
in nonlinear reactive dynamics at intermediate coupling ranges, whereas the Turing patterns 
emerge at the limit of local diffusive coupling. To that purpose, we study the generic Brusselator 
model \cite{nicolis:1977,lefever:1971}, which 
was proposed as a reduction of the complex chemical mechanism representing
 the Belouzov-Zhabotinsky reaction.
For this model we show, in a unifying description, that by appropriate choice of the
parameter values and coupling ranges we can pass from the spatial Turing patterns to the
oscillating spatiotemporal chimera states. In the simulations coupled
Brusselators linked in 2D torus
geometry are employed, where both Turing patterns and chimera states demonstrate
spectacular formations.
\end{quotation}

\section{Introduction}
\label{Introduction}

The motivation of the present study is to describe, in a unifying way, the emergence of
spatial structures and spatiotemporal patterns in different regions of the parameter space
in the same dynamical system. For this reason we use a typical nonlinear dynamical system
with minimal number of variables (two), the Brusselator, constructed specifically to model the
complex spatiotemporal organization of the Belouzov-Zhabotinsky reaction (see \cite{nicolis:1995} 
and references therein). For this model, we numerically demonstrate
 that it is possible to drive the system from spatial Turing
pattern to spatiotemporal oscillating chimera states
by controlling the range of the diffusion
from classical (local) diffusion to intermediate range (nonlocal) diffusion   with appropriate 
choice of the reaction rates (parameters). 

Turing patterns were first predicted in 1952 by Alan Turing in a  set of 
 nonlinear PDEs
describing a hypothetical chemical reaction, which involved autocatalytic steps
and species diffusion \cite{turing:1952}. 
Experimentally, these states were observed in several reactive systems beyond 
the Belouzov-Zhabotinsky, such as the Bray–Liebhafsky,
the Briggs–Rauscher and the Chlorite-Iodide-Malonic Acid (CIMA) reactions 
\cite{epstein:1998,epstein:1996,mikhailov:2006,winfree:1978,dekepper:1991,castets:1990}.
Regarding the numerical evidence of Turing patterns, until the 
late 1980's it was generally admitted in the literature that 
``Albeit their prediction by Turing, chemical periodic stationary patterns under
 far from equilibrium conditions, ... have not yet been observed in a clear-cut way''
\cite{borckmans:1987}. Nevertheless, 
a few years later, in the early 1990s, numerical simulations corroborated the
experimental findings and the theoretical predictions. Turing patterns were then reported 
 using a number of models,
including the Brusselator \cite{verdasca:1992,borckmans:1992,pena:2001}, the two- and three-variable
Oregonator models \cite{field:1972,field:1974,becker:1985,berenstein:2014,hildebrand:2001,epstein:1996}
and the two-variable Zhabotinsky-Buchholtz-Kiyatkin-Epstein (ZBKE) model 
\cite{zhabotinsky:1993,nkomo:2013}.

On the other end, chimera states have recently been the subject of intensive studies since their
first discovery by Kuramoto and Battogtokh in 2002 \cite{kuramoto:2002,kuramoto:2002a}.
The term ``chimera'', proposed by Abrams and Strogatz 
\cite{abrams:2004}, dominated two years after their first introduction  
and refers precisely to a hybrid state. Indeed, in networks composed of identical and
identically linked oscillators, as Kuramoto and Battogtokh demonstrated, a hybrid state is 
often established which consists of stable, coexisting, coherent and incoherent domains.
Originally, the chimera states were observed under nonlocal network connectivity, but
later works using local connectivity and non-identical oscillators have also reported
emerging chimera states.
In the first studies, the phase oscillator was used as the dynamics on the network
nodes \cite{bolotov:2016,goldschmidt:2019,bolotov:2020}, 
while most recent studies have demonstrated the presence of chimera states in a number of
nonlinear oscillators, such as 
the Hodgkin-Huxley (HH) \cite{sakaguchi:2006},
the FitzHugh-Nagumo (FHN) \cite{omelchenko:2013,omelchenko:2015,schmidt:2017,zakharova:2017,shepelev:2019}, 
the Hindmarsh-Rose \cite{hizanidis:2014},  
the Van der Pol \cite{ulonska:2016},  
the Stuart-Landau \cite{yeldesbay:2014,gjurchinovski:2017,tumash:2019}, the
Leaky Integrate-and-Fire (LIF) \cite{olmi:2015,tsigkri:2017,tsigkri:2018} 
and other oscillator networks. In the above studies the presence of chimera (multichimera) states
were reported with one (many) coherent and incoherent domains. The patterns and multiplicity
of the chimeras depend on 
the parameters governing both the nodal dynamics and the network coupling. 
\par The role that the network architecture
plays in the form of the chimera states has also been the subject of 
recent intense investigations. Most of the studies cited above
where performed in a 1D connectivity with periodic boundary condition (K-ring connectivity).
Considerably fewer studies concern connectivities in higher spatial dimensions.
Studies in 2D-square (torus) connectivity have revealed coherent (and incoherent) domains
of circular disc shapes, stripes, spots, scroll waves etc \cite{schmidt:2017,maistrenko:2017,laing:2017,oomelchenko:2019}. 
Similar studies in 3D-cubic (hypertorus)
connectivity produce coherent and incoherent domains of higher complexity, such as spheres,
cylinders, layers and combinations of these 
\cite{oomelchenko:2019,maistrenko:2015,kasimatis:2018,maistrenko:2017,maistrenko:2020,koulierakis:2020}.

\par Almost one decade after the first theoretical prediction of chimera states, experimental 
evidence of these states were reported in a number of 
mechanical, physical and chemical systems consisting of interacting oscillatory
units. Some examples reported in the literature include chimeras in 
experiments in optical systems \cite{hagerstrom:2012},
in electronic circuits \cite{gambuzza:2014},
in mechanics (systems of coupled mechanical oscillators) 
\cite{martens:2013,kapitaniak:2014,dudkowski:2016,manoj:2021},
in biomedicine \cite{cherry:2008,mormann:2000,mormann:2003a,santos:2017},
and in 
reaction diffusion systems \cite{tinsley:2012,wickramasinghe:2013,schmidt:2014}.
Beyond experimental findings in the laboratory, chimera states have been associated 
with the uni-hemispheric sleep in mammals and 
birds \cite{rattenborg:2000,rattenborg:2006,ramlow:2019},
 with the onset of epileptic seizures \cite{mormann:2000,mormann:2003a}
and other biomedical conditions \cite{andrzejak:2016,ruzzene:2019,bansal:2019,koulierakis:2020}.
For reviews on the recent theoretical and experimental 
developments in the field of chimera states, see references
\cite{panaggio:2015,schoell:2016,zheng:2016,majhi:2019,oomelchenko:2018,zakharova:2020,oomelchenko:2022}.

\par Regarding the emergence of chimera states in reaction-diffusion
systems, the first such reports date as early as 2012. 
In particular, photochemically coupled Belousov-Zhabotinsky micro-oscillators have been
shown to produce complex synchronization phenomena, such as spiral wave chimeras, 
spiral wave core splitting and phase cluster states 
\cite{tinsley:2012,nkomo:2013,taylor:2015,nkomo:2016,totz:2018,totz:2020}. 
Numerical simulations using the dimensionless
Zhabotinsky-Buchholtz-Kiyatkin-Epstein (ZBKE) model with a time delay in the coupling between
the oscillators corroborate the experimental findings \cite{nkomo:2016,totz:2018,totz:2020}. 
These studies indicate that chimera states are not only theoretical predictions but are
realized in chemical experimental setups. 
Additional evidence of chimera states in reaction-diffusion systems has been
 reported by the present author and collaborators
in a different numerical study drawing from population dynamics
and reactive systems in cyclic reactions \cite{hizanidis:2015}. In this study,
 the Lattice Limit Cycle (LLC) model was considered with 4-th order nonlinearities. Nonlinearly
coupled LLCs were shown to produce chimera states in the spiking regime, when the bifurcation 
parameter increases away from the Hopf bifurcation.  More recent studies investigating
the presence of Turing patterns and chimera states in networks of nonlinear oscillators
have been reported in the literature for the phase oscillator \cite{oomelchenko:2021} 
and for coupled SQUID models \cite{hizanidis:2020}.

\par As earlier stated, in the present study
 we consider a well known reaction diffusion scheme, the Brusselator,
 proposed as a minimal prototype scheme for demonstrating the effects of nonlinear 
interactions in chemistry. Its minimal dynamics involves two interacting variables with 
3rd order nonlinearities and  an autocatalytic step.
 Numerous studies in the past have shown that this model
gives rise to interesting Turing patterns in 1D, 2D and 3D
spatial geometries, when normal diffusion in the form of 
nearest neighbor interactions is included in the dynamics
and the parameter values are chosen in the Turing bifurcation regime
\cite{borckmans:1992,verdasca:1992,pena:2001,kokolnikov:2006,budroni:2016,dedecker:2013,tildi:2016,shoji:2007}. 
In the sequel, we extend gradually the range of interactions,
going from nearest neighbors (normal diffusion) to nonlocal, intermediate ranges 
where the chimera states prevail. In parallel, we determine 
the reaction rates and parameter ranges which trigger the appearance of 
complex synchronization phenomena and chimera states in two spatial dimensions.

\par The organization of the study is as follows: In the next section, \ref{sec:model},
 we introduce the
uncoupled (\ref{sec:uncoupledBXL}) and the coupled (\ref{sec:coupledBXL})
Brusselator model and discuss its mean field dynamics
and the conditions and parameters for the Hopf and the Turing bifurcations.
At the end of the same section we introduce some quantitative measures appropriate
for the description of Turing patterns and chimera states. In section \ref{sec:turing},
we show that by taking the diffusing limit in the coupling range we obtain
Turing structures of different morphologies.  In section \ref{sec:chimeras},
we increase the coupling range and for appropriate parameter values 
(after Hopf bifurcation) we demonstrate the emergence of chimera states.
By scanning the coupling range and coupling angle parameters we determine 
the region in the parameter space where the chimera states dominate.
In section \ref{sec:comments}, the nature of Turing and chimera patterns
are discussed using the Fourier transforms of the simulated evolution of the systems' variables.
In the concluding section, we recapitulate our main results and propose
open problems.

\section{The Model}
\label{sec:model}

\par In this section we first present the ODEs of the uncoupled Brusselator
model and discuss, in brief, the Hopf bifurcation scenario giving rise to
a limit cycle. Next, we present the nonlocally coupled Brusselator network and
discuss the emergence of a Turing bifurcation at the limit of local coupling.
In the last part of this section, we introduce some quantitative indices 
 for quantifying the spatial and temporal properties of the system.

\subsection{The uncoupled Brusselator Model}
\label{sec:uncoupledBXL}

\par As discussed in the Introduction, the Brusselator model was introduced in late 1960's
by R.~Lefever, G.~Nicolis and I.~Prigogine as a minimal, two-variable, autocatalytic model producing all 
spatiotemporal complexity of the Belouzov-Zhabotinsky oscillatory
 reaction \cite{nicolis:1977,lefever:1971}.
It consists of two ODEs of 3rd order, with two variables, $X$ and $Y$, and two
control parameters, $A$ and $B$, as follows:
\begin{subequations}
\begin{align}
\label{eq01a}
\frac{dX}{dt}&=A-(B+1)X+X^2 Y \\
\label{eq01b}
\frac{dY}{dt}&=BX-X^2 Y .
\end{align}
\label{eq01}
\end{subequations}
\noindent In Eqs. \eqref{eq01}, the variables $X$ and $Y$
take positive values since they were first constructed to represent chemical species concentrations,
while the parameters $A$ and $B$ represent the initial product concentrations and remain constant
(considered as sources). 
\par The Brusselator model, Eqs. \eqref{eq01}, has a single fixed point, $X_s$,
\begin{eqnarray}
(X_s,Y_s)=(A,B/A).
\label{eq02}
\end{eqnarray}
\par Stability analysis around $(X_s,Y_s)$ reveals that this fixed point changes its asymptotic stability
when the parameter $B$ attains its critical value $B_c$:
\begin{eqnarray}
B_c=A^2+1.
\label{eq03}
\end{eqnarray} 
Namely, for 
$ B< B_c=A^2+1$, $(X_s,Y_s)$ is a stable node or
stable focus and the phase space trajectories are attracted to the
fixed point. At $B=B_c$ the dynamics around the fixed point changes
 from asymptotic stability to instability 
and beyond this point a limit cycle is created \cite{nicolis:1995}. Because limit cycles are stable 
oscillators under reasonable perturbations, they are prerequisites for the emergence of the
complex synchronized states, chimera states, when many limit cycles are coupled in networks.
For this reason, in section~\ref{sec:chimeras} the parameter values $A=1.0$ and $B=2.5-3.5$ will be used 
(such that $ B > A^2+1$),
when the network composed by Brusselator units will be tuned to produce chimera states.
\par On the other hand, in the Brusselator model
the Turing structures are produced away from the regime of limit cycles
and under the Turing bifurcation conditions, where the coupling terms need to be taken into
account. For this reason, the discussion on the precise Brusselator parameter values 
which lead to the emergence of Turing structures is postponed till section~\ref{sec:turing}, 
after the coupling terms have been included.

\subsection{The coupled Brusselator Model}
\label{sec:coupledBXL}

\par In this section, spatial coupling is introduced between units in 2D, square lattice geometry.  
Earlier studies have discovered spectacular Turing patterns in 2D, such as dots and
stripes, and the aim of the present work is to clarify the different parameter domains in
the 2D Brusselator network
where the Turing patterns and the chimera states reside. 
On the square 
lattice network containing $N \times N$ units,
 the Brusselator located at position coordinates $(i,j)$ is 
described by the time
dependent variables $X_{ij}$ and $Y_{ij}$. The communication between the Brusselators 
is achieved by the coupling
matrix $C$ whose elements, $c_{ij:kl}$, 
represent the strength of the link between Brusselators at
positions $(i,j)$ and $(k,l)$ on the network, with $i,j,k,l \le N$. 
 Recent investigations
 of synchronization phenomena on complex networks have shown that coupling matrices
 with strong cross-coupling terms $(w_{XY},\> w_{YX})$ and both  excitatory (positive) 
and inhibitory (negative) terms 
are necessary (but not sufficient) for partial synchronization in the form of chimera states
\cite{omelchenko:2013,oomelchenko:2018}.
Under these conditions,  the 
nonlocally coupled Brusselator dynamics on the 2D square lattice is governed by the following equations:

\begin{widetext}
\begin{subequations}
\begin{align}
\frac{dX_{ij}}{dt}&=A-(B+1)X_{ij}+X_{ij}^2Y_{ij}
+ \sum_{(k,l)}c_{kl:ij} \left[ w_{XX}(X_{kl}-X_{ij})+w_{XY}(Y_{kl}-Y_{ij})\right] \\
\frac{dY_{ij}}{dt}&=BX_{ij}-X_{ij}^2Y_{ij}+ \sum_{(k,l)}c_{kl:ij}
 \left[ w_{YX}(X_{kl}-X_{ij})+w_{YY}(Y_{kl}-Y_{ij})\right]
\end{align}
\label{eq04}
\end{subequations}
\end{widetext}
\par In Eqs.~\ref{eq04}, we implicitly assume that all Brusselators are identical since they
have the same parameters $A$ and $B$. 
\par The coupling matrix $C $ is limited by
 the range $R$ and strength $\sigma$ 
of the nonlocal interactions. The elements $c_{ij:kl}$ representing the interactions
between Brusselators at positions $(i,j)$ and $(k,l)$ have the form:
\begin{widetext}
\begin{eqnarray}
c_{ij:kl}=\begin{cases}
\dfrac{\sigma} 
{ \displaystyle{(2R+1)^2-1}}, & \text{when} \>\>
 i-R \le k\le i+R, \>\>\> j-R\le l \le j+R   \\
\>\>\>\>\>\>\>\>\>\>\>\>\>\> 0, & \text{otherwise}.
\end{cases}
\label{eq05}
\end{eqnarray}
\end{widetext}
\par Namely, the Brusselator at position $(i,j)$ is coupled with all other elements located
only within a rectangle
of size \\ $(2R+1)\> {\rm X} \> (2R+1)$, whose center is at position $(i,j)$. Each nonzero 
link carries constant weight (coupling strength) $\sigma$, 
while the overall contribution is normalized
by the total number of nonzero links, which is $(2R+1)^2-1$. In Eqs.~\eqref{eq04} and \eqref{eq05},
periodic boundary conditions on the torus are considered and, therefore, all indices $i,j,k,l,$ 
are taken 
$\mod N$.

\par Apart for the coupling matrix $C$ between the nodes of the network,
a rotational matrix, $W$, is introduced to account for the cross-coupling terms. $W$ contains
both diagonal coupling, in the form  $(w_{XX},w_{YY})$ and 
cross coupling $(w_{XY},w_{YX})$ and is parameterized by a single parameter $\varphi$,
which is close to $\pi /2$ if cross-coupling is dominant \cite{omelchenko:2013}.
This coupling phase $\varphi$ is similar to the phase-lag parameter $\alpha$ of the
paradigmatic Kuramoto phase oscillator model, which is widely used to generically
describe coupled oscillator networks. The coupling phase is necessary for the
development of nontrivial partial synchronization patterns, as has been shown for the Kuramoto model 
\cite{oomelchenko:2010} and 
for the FHN model \cite{omelchenko:2013}.  
This rotational matrix $W$  is also used in 
Refs.~\cite{omelchenko:2015,schmidt:2017,ramlow:2019,mikhaylenko:2019} 
and has the form:

\begin{equation}
W=
\begin{pmatrix}
w_{XX} & w_{XY}\\
w_{YX} & w_{YY}
\end{pmatrix}
=\begin{pmatrix}
\cos{\varphi} & \sin{\varphi}\\
-\sin{\varphi} & \cos{\varphi}
\end{pmatrix}.
\label{eq06}
\end{equation}
 In the present study different values of the coupling range $R$
 and the coupling phase 
$0 \le \varphi \le 2\pi $ are examined to determine the domains in the $(\varphi , R )$ parameter 
intervals where the Turing patterns or the chimera states dominate.

\subsection{Quantitative measures}
\label{sec:quantitative}

\par The synchrony/asynchrony in the coupled Brusselator system
 can be verified using the mean phase velocity profile,
as in most studies where many oscillators are interlinked.
The mean phase velocity $\omega_{ij}$
accounts for the average number of cycles that oscillator at position $ (i,j)$
has performed in the time unit  \cite{omelchenko:2015,tsigkri:2018}. Let us call
$Q_{ij}$ the number of complete cycles
that oscillator$ (i,j)$ has performed during
a certain time interval $\Delta T$. Then $\omega_{ij}$ is defined as:
\begin{eqnarray}
\omega_{ij}=2\pi \frac{Q_{ij}}{\Delta T}=2\pi f_{ij}\>\>,
\label{eq07}
\end{eqnarray}
 where $f_{ij}$ is the frequency of the oscillator at position $(i,j)$. 
Because the definitions of frequency and mean phase
velocity differ by just a factor of $2\pi$, in the following the two expressions 
(``frequency'' and ``mean phase velocity'') are used interchangeably.
\par Chimera states are characterized by
a distribution of frequencies, while the coherent states are characterized by a single, common
frequency. Full incoherence may also be characterized by a single, common frequency, in some cases.
Let us denote by $\omega_{\rm min}=\min\{\omega_{ij}\}$, $1\le i,j \le N$, 
 the minimum mean phase velocity recorded in the
Brusselator network. Similarly, $\omega_{\rm max}=\max\{\omega_{ij}\}$, $1\le i,j \le N$, is the maximum mean phase velocity observed. 
The difference 
\begin{eqnarray}
\Delta \omega =\omega_{\rm max}-\omega_{\rm min},
\label{eq08}
\end{eqnarray}
 is a measure of the extent of
the mean phase velocity distribution in the system. We may assume 
that if  $\Delta \omega > 2*\pi/\Delta T$,
then the difference of the mean phase velocity between nodes can be considered significant 
and may be an indication of the establishment of a chimera state in the network. 
If $\Delta \omega < 2*\pi/\Delta T$, this might be considered as a one-cycle over- or under-estimation
and can be regarded as the error threshold.

\par The Kuramoto order parameter is another measure often used  to quantify coherence/incoherence
in the system \cite{kuramoto:2002,bick:2016}. 
Let us denote $\theta_{ij}$ the instantaneous phase of Brusselator at position $(i,j)$,
which is defined as $\theta_{ij}=\arctan{\left[ Y_{ij}/X_{ij}\right] }$. The Kuramoto order parameter $Z$
is defined globally, over the entire system as:
\begin{equation}
Z=\frac{1}{N^2}\displaystyle{\left| \sum_{i,j}^N \mathrm{e}^{+i\theta_{ij}}\right|}.
\label{eq09}
\end{equation}
In Eq.~\eqref{eq09}, if the system is synchronous and all phases are identical, then $Z=1$.
If all phases are random as in the asynchronous state, the various phases in the sum
of Eq.~\eqref{eq09} cancel out and $Z=0$. For intermediate cases, where synchronous and asynchronous
regions are present, $0 <Z<1$, and this is the domain where chimera states can be observed.
On the other hand, even in cases that the oscillators in the network have identical mean phase
 velocities but they operate under constant phase lags, they can be mistakenly considered as
chimera states, because $Z<1$. 
Therefore, both measures $\Delta \omega >0$ and $0<Z<1$ are necessary to quantify synchronization
in networks of coupled elements.

\par Other measures such as the spatial and/or temporal Fourier spectra can be used
to further verify the spatial and/or temporal order of the network nodes, respectively.

\section{Turing Structures in 2D}
\label{sec:turing}

\begin{figure}[h]
\includegraphics[width=0.48\textwidth,angle=0.0]{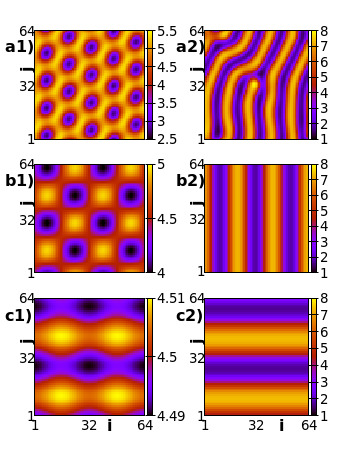}
\caption{\label{fig:01} Color-coded stable snapshots ($X-$variable profiles), Turing patterns,
of the coupled Brusselator network. 
{\bf a1)} Multiple spots for $B=6.75$ and $R=1$, {\bf a2)} multiple stripes, for $B=8$ and $R=1$, 
{\bf b1)} Multiple spots for $B=6.75$ and $R=2$, {\bf b2)} multiple stripes, for $B=8$ and $R=2$,
{\bf c1)} Multiple spots for $B=6.75$ and $R=3$ and {\bf c2)} multiple stripes, for $B=8$ and $R=3$.
For all integrations Eqs.~\eqref{eq04} and \eqref{eq05} were used, with the following parameters:
$N=64$, $A=4.5$, $D_X=7.0$, $D_Y=56$, $w_{XY}=w_{YX}=0 $, $h=1$, $\sigma =1$
with the integration constant $dt=0.001$. Images are recorded at 640 TUs.
Simulations start from random initial conditions with periodic boundary conditions.  
Two related videos are presented in the Supplementary Material 
(Multimedia view corresponding to images Fig.~\ref{fig:01}(a1)
 (movie\_fig1a1\_aa4.5\_bb6.75\_du7.0\_dv56.0\_dt0.001.mp4) and Fig.~\ref{fig:01}(a2) 
(movie\_fig1a2\_aa4.5\_bb8.00\_du7.0\_dv56.0\_dt0.001.mp4)).
}
\end{figure}

Classical Turing structures are obtained when a nonlinear reactive system (the Brusselator
in the present case) is subjected to interactions in the form of diffusion. The diffusion terms which
are added to the original Brusselator
model, Eqs.~\eqref{eq01}, involve the Laplacian operator and
have the form $D_X\nabla^2 X$ for Eq.~\eqref{eq01a} and 
$D_Y\nabla^2 Y$ for Eq.~\eqref{eq01b}. The parameters $D_X$ and $D_Y$ represent the diffusion coefficients
and Eqs.~\eqref{eq01} then become:
\begin{subequations}
\begin{align}
\label{eq10a}
\frac{dX}{dt}&=A-(B+1)X+X^2 Y + D_X\nabla^2 X \\
\label{eq10b}
\frac{dY}{dt}&=BX-X^2 Y + D_Y\nabla^2 Y .
\end{align}
\label{eq10}
\end{subequations}
The study of these equations 
demonstrates that the fixed point $(X_s,Y_s)$ undergoes a Turing bifurcation at a 
critical value $B^T_c$ of the parameter B, such that\cite{nicolis:1995,verdasca:1992,pena:2001}:
\begin{eqnarray}
B^T_c & = & (1+\eta A)^2, \nonumber \\  
\eta & = & \left( \frac{D_X}{D_Y}\right)^{1/2}<(1+A^2)^{1/2}-1 .
\label{eq11}
\end{eqnarray}
In addition, the discretization in space of the Laplacian operator retains only interactions with
the first neighbors, and Eqs.~\eqref{eq10} reduce to (see Appendix \ref{sec:appendix01}
 for the derivation):
\begin{subequations}
\begin{align}
\label{eq12a}
\frac{dX_{ij}}{dt}&=A-(B+1)X_{ij}+X_{ij}^2 Y_{ij}+\frac{D_X}{h^2}
\sum_{\{ k,l\} }\left[X_{kl}-X_{ij}\right] \\
\label{eq12b}
\frac{dY_{ij}}{dt}&=BX_{ij}-X_{ij}^2 Y_{ij} + \frac{D_Y}{h^2}
\sum_{\{ k,l\} }\left[Y_{kl}-2Y_{ij}\right].
\end{align}
\label{eq12}
\end{subequations}
Here the indices vary as $i-1 \le k \le i+1$ and $j-1 \le l \le j+1$, (i.e.,  the coupling range
$R=1$) and $h$ denotes the lattice constant. Equations ~\eqref{eq12}
 are of the same form as Eqs.~\eqref{eq04}, with the following parameter identifications  
\begin{itemize}
\item[a)] $ c_{ij;kl}\> w_{XX}=D_X/h^2$, 
\item[b)] $ c_{ij;kl} \> w_{YY}=D_Y/h^2 $,
\item[c)] $ w_{XY}=w_{YX}=0 $ and 
\item[d)] $R=1$.
\end{itemize}
With these parameter identifications and under restrictions Eq.~\eqref{eq11},
the generic coupled Brusselator system, Eqs.~\eqref{eq04}, may  produce Turing patterns as
shown in Fig.~\ref{fig:01}.
\par For the simulations in Fig.~\ref{fig:01} all systems started with the same random 
initial conditions. As working parameter set, generic values were used (obeying conditions Eq.~\eqref{eq11}),
namely,  $A=4.5$, $D_X=7.0$, $D_Y=56.0$, $w_{XY}=w_{YX}=0$. The constant $h=1$ 
as it is representing the lattice
constant (absolute distance between successive nodes), as described in the Appendix
\ref{sec:appendix01}.
On the left panels of Fig.~\ref{fig:01}
the parameter $B$ takes the value $B=6.75$, while on the right ones $B=8.0$.
All systems were integrated for 1500 Time Units (TUs), i.e., for $1500 \times 10^3$
integration steps. The Turing patterns stabilized between 300 and 400 TUs and remained
stable thereafter.
\par Different values of coupling range are used, $R=1$ at the top row, $R=2$ in the middle row
and $R=3$ in the last row, to demonstrate how the Turing patterns dissolve as
the coupling range increases from local (nearest neighbor)
coupling to nonlocal coupling. Note that the local coupling $R=1$ corresponds
directly to the diffusion operator, while for $R>1$ we step away from the dominance
 of the diffusion coefficient and the Turing bifurcation analysis is no longer valid.
\par In Fig.~\ref{fig:01} panel (a1), stable Turing spot patterns are recorded  
for $B=6.75$ and $R=1$ 
and all other parameters as in the working parameter set, described
in the previous paragraph and also provided in the legend of Fig.~\ref{fig:01}.
Note that the size of spots increases with $D_X$ and $D_Y$ (not shown).
In panel (a2) stable stripe patterns are recorded, for $B=8.0$ and $R=1$. In both 
panels (a1) and (a2) the parameters obey conditions Eqs.~\eqref{eq11} while 
the coupling range $R=1$,
attains the limit of diffusion on lattice  \cite{verdasca:1992,borckmans:1992,pena:2001}. 
\par The criterion on the coupling range is relaxed in the middle row of  Fig.~\ref{fig:01}.
Here the coupling range is increased to the value $R=2$ and now the size of the spots and stripes
increases leading to fewer spots of larger radius in Fig.~\ref{fig:01}(b1) 
and fewer  stripes of larger width in Fig.~\ref{fig:01}(b2).
Increasing further the coupling range to $R=3$ in the last row of Fig.~\ref{fig:01},
we note that the size of the spots increases
further and they merge into strips, see Fig.~\ref{fig:01}(c1). Note also that the range of the
$X-$ variable decreases inversely with $R$  and the system tends to the pure homogeneous 
state as $R$ increases. Similarly, in Fig.~\ref{fig:01}(c2) the width of the stripes
increases with $R$ and even fewer stripes are formed, leading again to the homogeneous 
state as $R$ increases further.
\par The reason for increasing of the size of the patterns (spots, stripes) with the
coupling range $R$ can be intuitively understood, because for larger $R$ each node
connects to a larger number of neighbors and therefore the elements can adjust their
variables at larger distances/areas. A similar effect can be also seen in the next section,
where the chimera states synchronize at larger distances/areas, as the coupling range
increases.

\par Specific features of the patterns in Fig.~\ref{fig:01} can be extracted via the Fourier transform, and the spatial wave length can be considered as a function of the coupling radius.
The results are shown in Table \ref{table01}.
\begin{table}
  \begin{tabular}{ | l | c | c || c | c | }
    \hline
     &   \multicolumn{2}{|c||}{Spots} &  \multicolumn{2}{|c|}{Stripes} \\
  \hline \hline
    R  & Amplitude & Wavelength & Amplitude & Wavelength \\ \hline
    1 & 1.03 & 15.84 & 2.7395(h) & 12.66(h) \\ \hline
    2 & 0.32 & 31.79 & 2.83(h) & 21.07(h)\\ \hline
    3 & 0.01 & 31.56 & 2.88(v) &31.62(v)\\\hline
    4 & 0.58 & 63.01 & 2.97(h) & 63.10(h)\\\hline
    5 & $7\times 10^{-8}$ & 63.01 & 2.44(h) & 63.10(h)\\    \hline
  \end{tabular}
\caption{The amplitude and wavelength of the spots and stripes patterns in Fig.~\ref{fig:01}
for different values of the coupling range $R$. The values related to the spot pattern
are averages over the $x-$ and $y-$ spatial directions. The notation (h) or (v) in the
stripes pattern denote that the amplitude and wavelength were computed in the horizontal
or vertical direction, respectively.
The parameter values are as in Fig.~\ref{fig:01}.}
\label{table01} 
\end{table}
\par In Table \ref{table01}, on the 2nd and 3rd columns,
the amplitudes and wavelengths of the spot pattern 
are presented for different $R$ values,
 as corresponding to Fig.~\ref{fig:01} (left column). They were computed as
averages over the amplitudes and wavelengths in the $x-$ and $y-$ directions.
 It is easy to understand
that as $R$ increases the wavelength increases and for $R=4$ and 5 one wavelength 
covers the entire system. In addition, for $R=4$ there is a non-negligible mode amplitude
 ($=0.58$), while for $R=5$ the amplitude tends to zero, indicating that the
system has reached the homogeneous steady state. 
\par Similar conclusions are reached 
for the stripe pattern. Depending on the orientation of the stripes,
 the amplitudes and wavelengths recorded in the 4th and 5th columns
of the Table \ref{table01}  were computed either in the 
horizontal ($x-$) direction 
and are denoted by the letter (h) or in the 
vertical ($y-$) direction 
and are denoted by the letter (v).
Here also, as $R$ increases from 1 to 2 and to 3, the wavelength doubles each time.
For $R=4$, one wavelength covers the entire system with non-negligible amplitude
($=2.97$). As the coupling range increases further to $R=5$, still one wavelength
covers the entire system, but the amplitude of the spatial oscillations decreases
($=2.44$). Therefore, the amplitude of the patterns decreases
with $R$, leading again to the homogeneous steady state for large $R-$values.
\par The results of Table~\ref{table01} and  Fig.~\ref{fig:01} highlight the special 
characteristics of the Turing patterns. They are best observed at the diffusion
limit, which corresponds to $R=1$ in the discrete lattice version of the equations,
and they dissipate as $R$ increases leading to homogeneous steady states. As we will
see in the next section, this is not the case for the chimera states. Not only
the chimeras are observed in the parameter regions
where limit cycles are supported but they also appear away from the diffusing
limit, at intermediate coupling ranges. 
\par We would like to stress here
 that we do not go into a deep study of the pattern formation of the
Brusselator model, this has been done extensively elsewhere 
\cite{borckmans:1992,verdasca:1992,pena:2001,kokolnikov:2006,budroni:2016,dedecker:2013,tildi:2016,shoji:2007}. We only
highlight  the dissolution of the Turing patterns as the coupling range increases,
in contradistinction with the behavior of chimera states, which are not supported
at the diffusing limit, but only at intermediate coupling ranges
(provided that all oscillators are identical).

\section{Chimera states in 2D}
\label{sec:chimeras}
\begin{figure}[b]
\includegraphics[width=0.48\textwidth,angle=0.0] {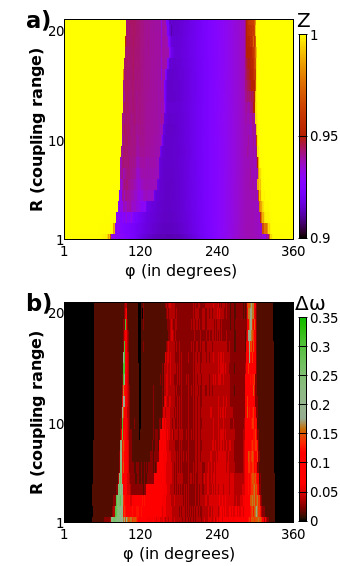}
\caption{\label{fig:02}  Color-coded values of a) the Kuramoto order parameter $Z$
and b) the frequency deviation in the system, $\Delta\omega$, as a
function of the coupling range $R$ and the coupling angle $\varphi$. 
Other parameters for the Brusselator model are $A=1.0$, $B=2.9$. 
Integration parameters are set to $dt=0.010$, $D_X=D_Y=1$. The network
is a 2D square lattice with periodic boundary conditions and
size $N \times N=64 \times 64$. All simulations start from the same
random initial conditions.
}
\end{figure}
The same model, Eqs.~\eqref{eq04} with connectivities as in ~\eqref{eq05} and ~\eqref{eq06}, 
for appropriate parameter values and intermediate coupling ranges
can lead to oscillatory steady states with coexisting variable frequencies
on the network, known as chimera states. To demonstrate this, we integrate 
numerically the afore mentioned equations, using the following working
parameter set: $A=1.0$, $B=2.9$ to place the uncoupled Brusselator in the parameter
domain where limit cycles are observed, see section~\ref{sec:uncoupledBXL}.
The integration parameters are set to $dt=0.010$, $D_X=D_Y=1$ and the 2D network
consists of a square lattice composed by $N \times N=64 \times 64$ elements,
with periodic boundary conditions (same as in the previous section, \ref{sec:turing}, 
where the Turing patterns are produced). The coupling 
strength $\sigma$ is set to unity. To explore the presence of chimera states
in the Brusselator network, the coupling range $R$ 
and coupling angle $\varphi$ were varied in the ranges, 
$1 \le R \le 21$ and $0\le \varphi < 2\pi$ (or $0\le \varphi < 360$ in degrees).

\par As discussed in section~\ref{sec:quantitative}, the presence of chimera states
is characterized by values of the Kuramoto order parameter $Z$ such that $0 < Z <1$
and $\Delta\omega >0$. In Fig.~\ref{fig:02}a, we present the values of $Z$
in a color-coded map, as a function of $R$ and $\varphi$. We note that chimera states
can be sought  in the regions that the $Z$-values take dark colors and away from the
yellow regions. Similarly, in Fig.~\ref{fig:02}b, we present the values of 
$\Delta \omega$ in a color-coded map. In the $\Delta \omega$ map, there is evidence
of chimera presence in the red and green regions which support high divergence
of local frequencies in the system. 

\par Both images, Fig.~\ref{fig:02}a and b, give compatible information as far as
the parameter regions where chimera states are realizable. Namely,
for small values of $\varphi <~90^{\circ}$ and all values of $R$ chimera states
are not possible: the Kuramoto order parameter $Z \sim 1$, the
frequency distribution $\Delta\omega \sim 0$ then coherence reins in the system. 
Similarly, chimera states are
not promoted for high values of $\varphi > 280 ^{\circ}$. 
This is in agreement with previous studies
of other models (such as the FitzHugh Nagumo model) which study and
report chimera states for $\varphi$ values away from 0 or $2\pi $.\cite{omelchenko:2013,omelchenko:2015}. 
Although there is a slight dependence on the coupling range $R$, 
as both indices $Z$ and $\Delta \omega$ indicate, this dependence is
not very strong. Comparing the two maps, the $\Delta\omega$ one (Fig.~\ref{fig:02}b)
shows higher details in the structure and higher ramifications in the frequency
distributions.

\par
\begin{figure}[h]
\includegraphics[width=0.50\textwidth,angle=0.0] {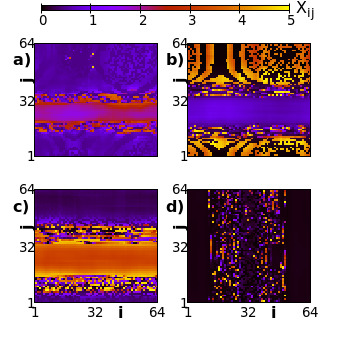}
\caption{\label{fig:03} 
Synchronous and asynchronous stripes for low $\varphi$-values. Spatial $X-$profiles
 for: a) $(\varphi =91^{\circ},\>\> R=9)$,  b) $(\varphi =93^{\circ},\>\> R=14)$,  
  c) $(\varphi =96^{\circ},\>\> R=18)$ and d) $(\varphi =97^{\circ},\>\> R=19)$.
Other parameters are $A=1.0$, $B=2.9$, $dt=0.010$, $D_X=D_Y=1$, $N=64 $ 
and $\sigma =1$. Toroidal periodic boundary conditions are applied. 
The images are recorded at 1000 TUs.
All simulations start from the same
random initial conditions. A related video is presented in the Supplementary Material 
(Multimedia view corresponding to image Fig.~\ref{fig:03}c
 (movie\_fig03c\_brusselator\_sig\_1.00\_phi\_096\_R\_18.mp4)).
}
\end{figure}

 Our numerical investigations demonstrate that interesting chimera states are produced
when the values of the coherence indices escape the $(Z,\Delta\omega ) \sim (1,0)$ values. 
In particular, at the borderlines between the yellow (black-grey)
 and the dark regions (red-green) in Fig.~\ref{fig:02}a (b), one may observe the formation
of stripe chimeras, several examples of which are provided in Fig.~\ref{fig:03}.
In this figure, where $\varphi \approx 90^{\circ}$ and different values of $R$ are used, alternation
of coherent and incoherent stripes is recorded. Judging from the color-coded maps,
Figs.~\ref{fig:02}a,b,
we note that
these chimera patterns are displayed for borderline parameter values, where
$\Delta\omega$ just exceeds 0 and  $Z$ just falls below 1.
In Figs.~\ref{fig:03}a and c two synchronous and 2 asynchronous stripes develop,
while in panels b and d only one synchronous and one asynchronous stripe are recorded.
It is worth to note that in  Fig.~\ref{fig:03}a, b, c the stripes develop horizontally,
while in d the stripes are vertical. This is due to the evolution of the 
random initial conditions, which are common in all simulations. Other pairs with notable
stripe chimeras are recorded for $(\varphi , R )$ values such as 
$(91^{\circ}, 8)$, $(92^{\circ}, 9)$,
$(92^{\circ},10)$, $(94^{\circ}, 14)$, $(94^{\circ}, 15)$, 
$(95^{\circ}, 15)$, $(96^{\circ},15)$, $(95^{\circ}, 16)$, 
$(96^{\circ},16)$, $(95^{\circ},17)$, $(96^{\circ},17)$ and others.

\par One significant difference can be stressed between images in Fig.~\ref{fig:01} (Turing patterns)
and those in Fig.~\ref{fig:03} (chimera states) and the ones that follow. In  Fig.~\ref{fig:01}
the images are stationary (still) after transient and do not change with time, 
while in Fig.~\ref{fig:03}
(and lateron) the images are dynamic: the colors (state variables) oscillate in time,
but the synchronous areas remain synchronous and so do the asynchronous ones. This means
that in the color-coded spacetime plots in Fig.~\ref{fig:03}, the colors of all nodes (pixels) 
change continuously between black and yellow. To the contrary,
 in Fig.~\ref{fig:01} they remain fixed after an initial transient period.

\par
\begin{figure}[h]
\includegraphics[width=0.50\textwidth,angle=0.0] {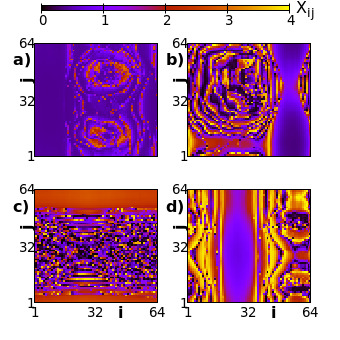}
\caption{\label{fig:04} 
Synchronous and asynchronous stripes for high $\varphi$-values. Spatial $X-$profiles for:
a) $(\varphi =298^{\circ},\>\> R=15)$, b) $(\varphi =299^{\circ},\>\> R=14)$, 
c) $(\varphi =295^{\circ},\> R=20)$ and d) $(\varphi =300^{\circ},\> R=19)$.
  Other parameters as in Fig.~\ref{fig:02}.
All simulations start from the same
random initial conditions. A related video is presented in the Supplementary Material 
(Multimedia view corresponding to image Fig.~\ref{fig:04}c
 (movie\_fig04c\_brusselator\_sig\_1.00\_phi\_295\_R\_20.mp4)).
}
\end{figure}

 Stripes are also recorded when the high values of $\Delta\omega$ (low values of $Z$) cross
to lower values of $\Delta\omega$ (high values of $Z$); this takes place
 around $\varphi \approx 300^{\circ} $, for different $R$-values. Spatial portraits
of typical stripe chimeras in these parameter region are
reported in Fig.~\ref{fig:04}. Panel a shows coexistence of a coherent (blue) and an 
incoherent stripe. In panel b, the coherent stripe has variable width. Such states are
often unstable and they result in a turbulent incoherent state.  Panels c (horizontal stripes)
and d (vertical stripes) are stable chimeras, recorded for higher $R$ values.
Other parameter values producing interesting stripe chimeras in this area are the pairs
$(\varphi ,\>\> R )$:  
$(300^{\circ}, \>\> 14)$, $(299^{\circ}, \>\> 15))$, $(295^{\circ}, \>\> 16)$,
$(298^{\circ}, \>\> 19)$, $(299^{\circ}, \>\> 19)$,
$(296^{\circ}, \>\> 20)$, $(297^{\circ}, \>\> 20)$,  $(298^{\circ}, \>\> 20)$ 
and $(299^{\circ}, \>\> 20)$.

\par We can also  observe transitions from coherent traveling linear fronts to traveling 
chimera cores via the production of solitary states for rotation angle values around
 $\varphi \approx 220^{\circ}-240^{\circ}$, in the center of the parameter regions of Fig.~\ref{fig:02}.
In this parameter region
$Z<1$ and $\Delta\omega >0$. Keeping $R=21$, we gradually increase $\varphi$ and
we present the transitions from traveling coherent fronts to traveling coherent/incoherent
domains, in Fig.~\ref{fig:04}. In this figure, the spatial $X-$profiles for $\varphi =220^{\circ},\> R=21$ in
panel a show the development of coherent traveling fronts. As the rotation coupling angle
increases to $\varphi = 225^{\circ}$ we note in panel b the presence of a few (three)
solitary Brusselators whose state variables escape from those of the neighbors.
When $\varphi $ is further increased to $\varphi = 228^{\circ}$, more solitaries mobilize causing
perturbations in the linear structure of the traveling fronts, see panel c. For 
$\varphi = 229^{\circ}$ the solitaries organize into incoherent regions as shown in panel d.
Further increase of $\varphi$ to $\varphi =236^{\circ}$ causes increase in the size of the 
incoherent regions and
total deformation of the linear structure of the traveling fronts. Finally, for
$\varphi=236^{\circ}$ the coherent fronts break totally and incoherent cores are formed as satellites
to traveling coherent domains. 
Indeed, one can observe ramifications in the $\Delta\omega$ values in
the central part of Fig.~\ref{fig:02}b, which justify the presence of traveling front
(coherent structures) with $\Delta\omega =0$ (black regions) next to traveling chimera
states with $\Delta\omega >0$ (red and blue regions). Similar behavior involving 
transitions from coherent traveling fronts to moving chimera states via solitary mobilization
is observed for large and intermediate values of $R$ and intermediate $\varphi$ domains.

\begin{figure}[h]
\includegraphics[width=0.50\textwidth,angle=0.0] {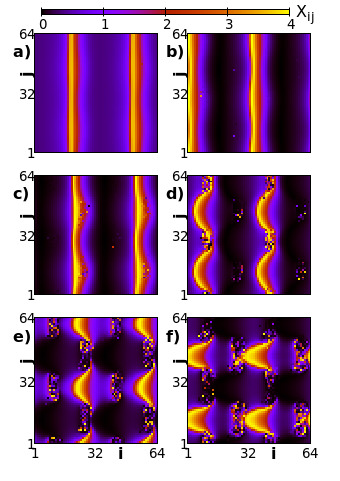}
\caption{\label{fig:05} 
Transitions from coherent stripes to solitaries and further to coherent and
incoherent traveling domains. Spatial $X-$profiles for: 
a) $(\varphi =220^{\circ}, \>\> R=21)$, b) $(\varphi =225^{\circ}, \>\> R=21)$, 
c) $(\varphi =228^{\circ}, \>\> R=21)$, d) $(\varphi =229^{\circ}, \>\> R=21)$,
e) $(\varphi =230^{\circ}, \>\> R=21)$ and f) $(\varphi =236^{\circ}, \>\> R=21)$. 
Other parameters as in Fig.~\ref{fig:02}.
All simulations start from the same
random initial conditions. A related video is presented in the Supplementary Material 
(Multimedia view corresponding to image Fig.~\ref{fig:05}e
 (movie\_fig05e\_brusselator\_sig\_1.00\_phi\_230\_R\_21.mp4)).
}
\end{figure}

\par
\begin{figure}[h]
\includegraphics[width=0.50\textwidth,angle=0.0] {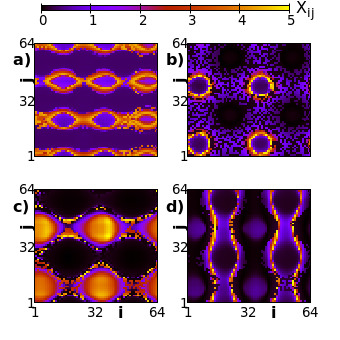}
\caption{\label{fig:06} 
Multiple connected and disconnected coherent spots. Spatial 2D $X-$profiles for: 
a) $(\varphi =160^{\circ}, \>\> R=16)$, b) $(\varphi =160^{\circ}, \>\> R=21)$, 
c) $(\varphi =165^{\circ}, \>\> R=19)$ and d) $(\varphi =166^{\circ}, \>\> R=19)$.
 Other parameters as in Fig.~\ref{fig:02}.
All simulations start from the same
random initial conditions.
}
\end{figure}

 In the vicinity of  $\varphi \approx \pi(=180^{\circ})$ 
we can observe multiple coherent spots surrounded
by an incoherent halos. Depending on the parameter 
value, the size and number of the coherent spots varies to cover exhaustively the 2D lattice. Representative 
examples are provided in Fig.~\ref{fig:06}. In panel a, for $\varphi =160^{\circ},\>\> R=16$,
three zones of interconnected coherent spots are observed. These zones are separated
by coherent ones of varying width. The borders between the coherent spots and the
coherent zones consist of incoherent elements. The coherent spots and coherent 
zones are characterized by a small, but persistent phase difference.
When the connectivity range increases to $R=21$
in panel b,
the number (and form) of the coherent spots decreases (from 9 to 4) and the area
of the incoherent domains grows. The coherent strings also degenerate into 4 coherent
domains lagging behind with respect to the value of the state variable as compared
to the coherent spots. 
 In panel c, for different parameter values (in the same parameter
domain) $\varphi =165^{\circ},\>\> R=19$, the size of the connected coherent spots increases and
their multiplicity decreases to 4. Here the phase difference between the coherent spots
and the coherent zones is considerable, the former being in the red-yellow colors
(highest $X$-values), while the latters are in the black-blue colors (lowest $X$-values). By increasing slightly $\varphi$ to $166^{\circ}$, in panel d, the phase difference between the coherent spots
and the coherent zones vanishes and the connecting incoherent domains are now visible
in vivid colors.

\begin{figure}[h]
\includegraphics[width=0.50\textwidth,angle=0.0] {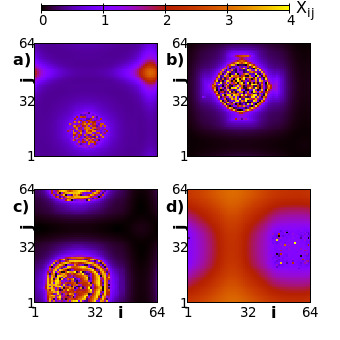}
\caption{\label{fig:07} 
Single incoherent spots. Spatial 2D $X-$profiles for: 
a) $(\varphi =303^{\circ}, \>\> R=6)$, b) $(\varphi =302^{\circ}, \>\> R=10)$, 
c) $(\varphi =304^{\circ}, \>\> R=8)$ and d) $(\varphi =302^{\circ}, \>\> R=18)$.
 Other parameters as in Fig.~\ref{fig:02}.
All simulations start from the same
random initial conditions. Two related videos are presented in the Supplementary Material 
(Multimedia view corresponding to image Fig.~\ref{fig:07}b
 (movie\_fig07b\_brusselator\_sig\_1.00\_phi\_302\_R\_10.mp4) and Fig.~\ref{fig:07}c
(movie\_fig07c\_brusselator\_sig\_1.00\_phi\_304\_R\_08.mp4)).
}
\end{figure}

Along the right border between turbulent states and coherence, 
stable incoherent spots may develop for appropriate values of the system parameters. These
can be considered as shrinking or degeneration of the stripes which are observed in nearby parameter areas (see Fig.~\ref{fig:04}).
In Fig.~\ref{fig:07} a variety of incoherent spots are recorded, for parameter values
$302 \le \varphi \le 304$ and different $R$.

Besides the  patterns discussed  above, we also report the following ones:
\begin{itemize}
\item For $\varphi \approx 170^{\circ}-200^{\circ}$, traveling square coherent domains develop.
Depending on the values of $R$ the coherent squares can be surrounded by incoherent halos.
\item Linear coherent stripes cross the system for large values of $R$, e.g., $R=21$
and $203^{\circ} \le \varphi \le 225^{\circ}$ and other parameters values in the same area.
\item Combinations of coherent linear parts and incoherent irregular shapes
are often recorded. These are observed for smaller values of $R$, e.g., $R=6$ and similar
values.
\item Scroll waves are recorded for parameter values near the stable chimera patterns, 
when the incoherent domains dissolve, e.g.,  for $\varphi \approx 300^{\circ}$ and $2 \le R \le 10$.
 Examples of these and other
turbulent chimera patterns are shown in Appendix \ref{sec:appendix02}.
\item Results for the case $R=1$, which corresponds to the diffusion limit, are also
shown in Appendix \ref{sec:appendix02}.
\end{itemize}

\par The parameter domain were the various patterns dominate were obtained for
 the working parameter set given in the first
paragraph of the present section. For different generic parameter values
the borders between the various domains of coherence are adjusted appropriately.

\section{General comments on spatial and spatiotemporal pattern formation 
in the Brusselator model}
\label{sec:comments}
\par From the results in the previous two sections, it becomes clear that in
different parameter domains the Brusselator network exhibits different spatial
(Turing) and spatiotemporal (chimera) patterns of high complexity. 
\par
\begin{figure}[h]
\includegraphics[width=0.40\textwidth,angle=0.0]{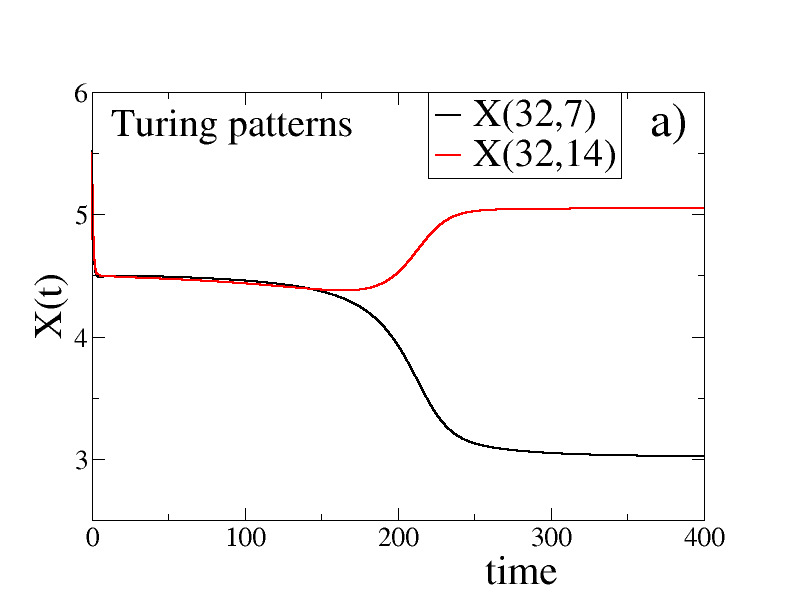}
\includegraphics[width=0.40\textwidth,angle=0.0]{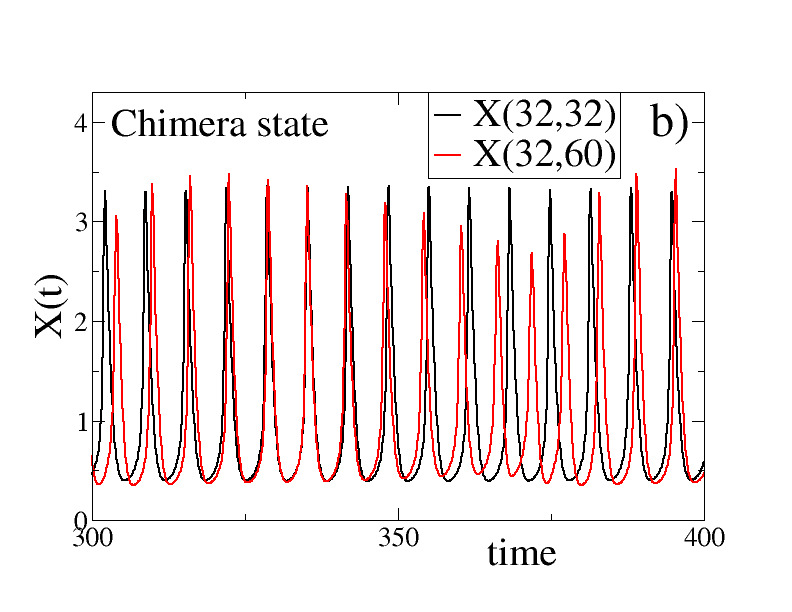}
\caption{\label{fig:08} 
Temporal evolution of Brusselator elements: 
a) For parameter values in the Turing regime, as in Fig.~\ref{fig:01}a1. The
black solid (lower) line corresponds to the element $(32,7)$ in the center of a
spot and  the red (upper) line to the element $(32,14)$ outside the
spot areas.  Other parameters as in Fig. \ref{fig:01}a1.
b)  For parameter values in the chimera regime, 
as in Fig.~\ref{fig:04}c. The
red solid line corresponds to the element $(32,60)$ in the coherent
region and the black line to the element $(32,32)$ in the incoherent stripe.
 Other parameters as in Fig. \ref{fig:04}c. 
}
\end{figure}

\begin{figure}[ht]
\includegraphics[width=0.48\textwidth,angle=0.0]{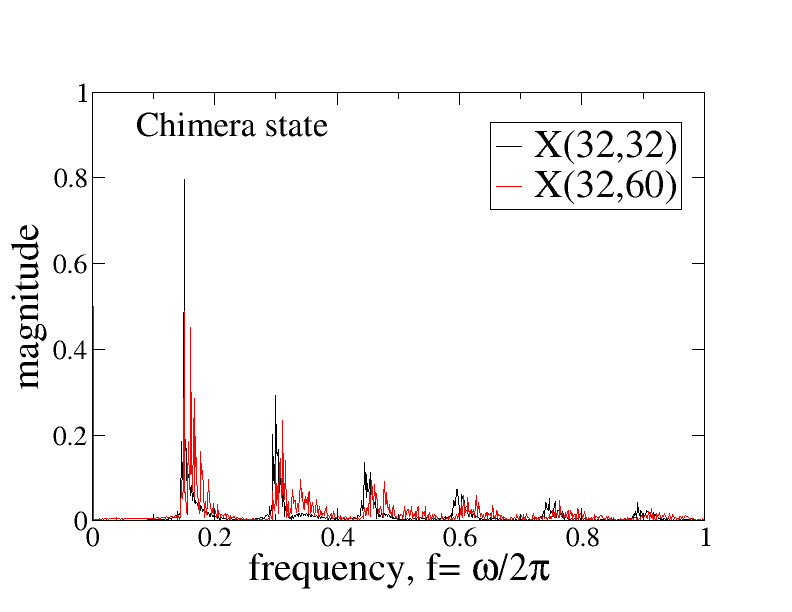}
\caption{\label{fig:09} 
Fourier spectra corresponding to the temporal evolution of
the Brusselator elements shown in Fig.~\ref{fig:08}b. 
The
parameter values and line descriptions follow the ones in Fig. \ref{fig:08}b.
For the extraction of the
Fourier spectra DFT Fourier transform was used and time series covered 
approximately approximately 170 cycles, in both cases, excluding the
first 50 cycles as transient.
}
\end{figure}

 Indeed, Turing patterns describe in section~\ref{sec:turing} are spatial patterns
which are stable in time after a transient period, while in chimera states
all elements perform nonlinear oscillations, but they are spatially organized 
in coherent and incoherent domains.
As an additional evidence of the intrinsic difference in the
temporal evolution between these two types of patterns, we present
in the two figures below the temporal evolution of particular oscillators, in Fig.~\ref{fig:08},
 and their corresponding
Fourier spectra,  in Fig.~\ref{fig:09}. 
We have chosen to demonstrate this for parameters as in Fig.~\ref{fig:01}a1
 corresponding to the Turing regime and Fig.~\ref{fig:04}c in the chimera regime.
\par From Fig.~\ref{fig:08}a,
it is evident that the Brusselators with parameters set at the Turing regime
 reach a stable steady state, different
for each element. Namely, the $X-$variable of Brusselator at position (32,7), which corresponds
to the center of a spot in Fig.~\ref{fig:01}a1, reaches value $ X(32,7)\to \sim 3$,
while the one at position (32,14), which is located outside the spot areas,
 reaches $ X(32,14)\to \sim 5$. To the contrary, when the Brusselators are
set in the chimera regime, they all oscillate in time, as shown in Fig.~\ref{fig:08}b.
In Fig.~\ref{fig:08}b (red line), we have chosen to demonstrate the temporal evolution 
of a node  in the coherent domain, (node (32,60) in Fig.~\ref{fig:04}c).
In the same figure the black line shows the evolution of  
a node  in the 
incoherent domain (node (32,32) in Fig.~\ref{fig:04}c). 
Both elements show similar oscillations, however, we may observe
that the Brusselator which belongs to the incoherent region, position (32,32),
is characterized by lower frequency of oscillations than the one at position
(32,60) in the coherent region. This is a well known property of chimera
states, where the coherent and incoherent regions are characterized by
different frequencies\cite{kuramoto:2002,omelchenko:2015,schoell:2016,oomelchenko:2018}, 
as will be also verified in the next figure.  
Similar results and
conclusions are obtained for all other parameters in the chimera regime.

\par To investigate the difference in the frequencies in the coherent and
incoherent domains of the chimera states, as an example we plot in Fig.~\ref{fig:09}
the Fourier spectra of the two time series shown in Fig.~\ref{fig:08}b.
We choose again the same two Brusselator elements at positions (32,60), in the coherent region,
plotted with the red line and at position (32,32), in the incoherent region, plotted 
with black line. From the plot, it is evident that for both elements only one frequency prevails,
together with the corresponding harmonics. The maxima in the two curves are at
different positions, consistently for the main frequencies and their harmonics.
This observation corroborates the statement in the previous paragraph on the difference in the
oscillator frequencies in  the coherent and the incoherent regions of the chimera states.

\section{Conclusions}
\label{sec:conclusions}

\par We have investigated the conditions under which Turing patterns and chimera states
can be produced in the same nonlinear dynamical system. Using as exemplary case the Brusselator model,
we have demonstrated Turing patterns for parameter regions where the single node dynamics presents
fixed points. 
The same dynamical system exhibits chimera states if the parameters of the
 single element  are chosen in the oscillatory (limit cycle) regime. Connectivity-wise,
for the emergence of the Turing patterns small connectivity neighborhoods are required, mimicking
 the nearest neighbor, Laplacian diffusion, while for the production of chimera states the 
connectivity regions need to be of intermediate range.

\par Based on the above results and discussions, 
we may add that it is not possible to obtain chimera states
and Turing patterns for the same values of parameters $A$ and $B$ in the Brusselator model,
 evenif the other (connectivity) parameters are different.
This is because for the emergence of Turing patterns (stable steady states)
the single element dynamics need to lead to
fixed points, while for the emergence of chimera states (oscillatory steady states)
a limit cycle is necessary in the 
single element dynamics. The above conclusions
hold provided that the coupling terms do not drastically alter 
the nodal dynamics. Additional numerical  and analytical studies using
 different coupled nonlinear
models are necessary to further elucidate the necessary and sufficient conditions for the
presence of chimera states in these systems.


\begin{acknowledgments}
The author (A.P.) acknowledges helpful discussions with K. Anesiadis.  
This work was supported by computational time granted from the 
Greek Research \& Technology Network (GRNET) in the National HPC facility - ARIS, 
 under project IDs PR009012 and PR12015.

\end{acknowledgments}

\section*{Data Availability Statement}
The data that support the findings of this study are available from the corresponding author upon reasonable request.

\appendix

\section{Discrete form of Brusselator with diffusion}
\label{sec:appendix01}
In this appendix we discretize the Brusselator model with diffusion and bring
it to the general
form, Eq.~\eqref{eq04}, in order to use a generic system of differential equations, 
both for the production
of Turing patterns and chimera states (at different regions of the parameter space).
\par Starting from Eq.~\eqref{eq10}, we specify the location $(i,j)$ of a Brusselator 
node and analyze the
Laplacian operator around this location. Then, Eq.~\eqref{eq10} reduces to:
\begin{subequations}
\begin{align}
\label{eq-appendix01a}
\frac{dX_{ij}}{dt}&=A-(B+1)X_{ij}+X_{ij}^2 Y_{ij} + D_X\nabla^2 X_{ij} \\
\label{eq-appendix01b}
\frac{dY_{ij}}{dt}&=BX_{ij}-X_{ij}^2 Y_{ij} + D_Y\nabla^2 Y_{ij} .
\end{align}
\label{eq-appendix01}
\end{subequations}
where $\nabla^2$ is the Laplacian operator in two dimensions and $D_X$, $D_Y$
are the diffusion constants of the $X$ and $Y$ variables, respectively. By expanding the Laplacian
operator in space around position $(i,j)$ we obtain:
\begin{widetext}
\begin{subequations}
\begin{align}
\label{eq-appendix02a}
\frac{dX_{ij}}{dt}&=A-(B+1)X_{ij}+X_{ij}^2 Y_{ij}+\frac{D_X}{(\Delta x)^2}
\left[X_{i+1,j}-2X_{ij}+X_{i-1,j}\right] + \frac{D_X}{(\Delta y)^2}
\left[X_{i,j+1}-2X_{ij}+X_{i,j-1}\right] \\
\label{eq-appendix02b}
\frac{dY_{ij}}{dt}&=BX_{ij}-X_{ij}^2 Y_{ij} + \frac{D_Y}{(\Delta x)^2}
\left[Y_{i+1,j}-2Y_{ij}+Y_{i-1,j}\right] + \frac{D_Y}{(\Delta y)^2}
\left[Y_{i,j+1}-2Y_{ij}+Y_{i,j-1}\right].
\end{align}
\label{eq-appendix02}
\end{subequations}
\end{widetext}
\par In Eqs.~\eqref{eq-appendix02}, $\Delta x$ and $\Delta y$ represent the infinitesimal displacement
on the x- and y-axis, respectively, as per the derivative definition. In the case of discretization,
the smallest possible displacement on the lattice is one lattice site, 
and therefore $\Delta x =\Delta y =h=1$
in the network discretization approximation. Seeing from a different point of view, the displacement values 
$\Delta x$ and $\Delta y$ can be absorbed in the factor $D_X$ and $D_Y$. Gathering together the terms
which are linked to the neighboring nodes, Eq.~\eqref{eq-appendix02} reduce further to:
\begin{widetext}
\begin{subequations}
\begin{align}
\label{eq-appendix03a}
\frac{dX_{ij}}{dt}&=A-(B+1)X_{ij}+X_{ij}^2 Y_{ij}+\frac{D_X}{(\Delta x)^2}
\sum_{k=i-1}^{i+1}\left[X_{kj}-X_{ij}\right] + \frac{D_X}{(\Delta y)^2}
\sum_{l=j-1}^{j+1}\left[X_{il}-X_{ij}\right] \\
\label{eq-appendix03b}
\frac{dY_{ij}}{dt}&=BX_{ij}-X_{ij}^2 Y_{ij} + \frac{D_Y}{(\Delta x)^2}
\sum_{k=i-1}^{i+1}\left[Y_{kj}-Y_{ij}\right] + \frac{D_Y}{(\Delta y)^2}
\sum_{l=j-1}^{j+1}\left[Y_{il}-Y_{ij}\right].
\end{align}
\label{eq-appendix03}
\end{subequations}
\end{widetext}
\centerline{}

\centerline{}

\centerline{}

\centerline{}

\centerline{}
\vskip 2cm
\centerline{}

\centerline{}
\par Finally, gathering the terms on the right-hand-sides and because $\Delta x =\Delta y =h=1$,
 Eqs.~\eqref{eq-appendix03} result in:
\begin{subequations}
\begin{align}
\label{eq-appendix04a}
\frac{dX_{ij}}{dt}&=A-(B+1)X_{ij}+X_{ij}^2 Y_{ij}+\frac{D_X}{h^2}
\sum_{\{ k,l\} }\left[X_{kl}-X_{ij}\right] \\
\label{eq-appendix04b}
\frac{dY_{ij}}{dt}&=BX_{ij}-X_{ij}^2 Y_{ij} + \frac{D_Y}{h^2}
\sum_{\{ k,l\} }\left[Y_{kl}-Y_{ij}\right].
\end{align}
\label{eq-appendix04}
\end{subequations}
 Equations ~\eqref{eq-appendix04} have the same form with the more general Eqs.~\eqref{eq04},
with the following restrictions:
\begin{subequations}
\begin{align}
\label{eq-appendix05a}
& c_{ij;kl}\> w_{XX} =D_X/h^2, \\
\label{eq-appendix05b}
& c_{ij;kl}\> w_{YY} =D_Y/h^2,\\
\label{eq-appendix05c}
& w_{XY}=w_{YX}=0,\\
\label{eq-appendix05d}
& R=1.
\end{align}
\label{eq-appendix05}
\end{subequations} 
I.e., the more generic form of the Brusselator network, Eqs. ~\eqref{eq04}, can give rise to 
Turing patterns when cross-coupling terms are absent from the dynamics and in the limit of
local, nearest-neighbor exchanges, i.e., under the conditions Eqs.~\eqref{eq-appendix05}.

\section{Additional turbulent chimeric patterns}
\label{sec:appendix02}

\par In the present section additional interesting patterns are recorded,
composed by coexisting coherent and
incoherent traveling patterns. 
\par
\begin{figure}[h]
\includegraphics[width=0.50\textwidth,angle=0.0] {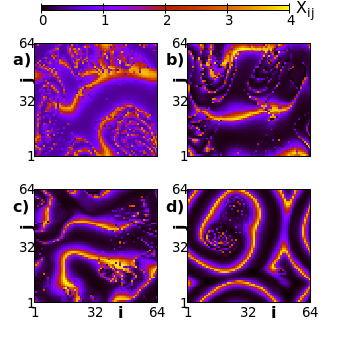}
\caption{\label{fig:turbulent-patterns} 
Turbulent patterns. Spatial 2D $X-$profiles for: 
a) $(\varphi =303^{\circ}, \>\> R=6)$, b) $(\varphi =302^{\circ}, \>\> R=10)$, 
c) $(\varphi =304^{\circ}, \>\> R=8)$ and d) $(\varphi =302^{\circ}, \>\> R=18)$.
 Other parameters as in Fig.~\ref{fig:02}.
All simulations start from the same
random initial conditions.
}
\end{figure}

 First, in Fig.~\ref{fig:turbulent-patterns} turbulent
traveling patterns are shown. Of particular interest is panel d, which consists of traveling
fronts (red colors). One of the fronts forms a spiral pattern at the interior of which 
an incoherent core dominates.  Similar incoherent cores are reported in Refs. 
\cite{totz:2018,totz:2020} where the ZBKE model is used.

\par
\begin{figure}[h]
\includegraphics[width=0.45\textwidth,angle=0.0] {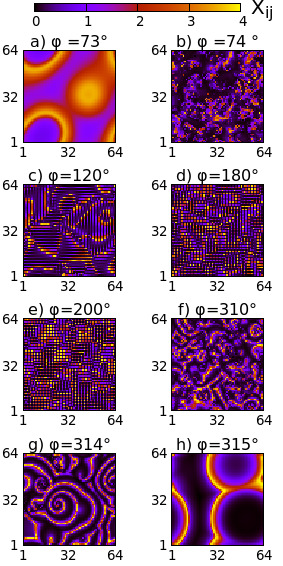}
\caption{\label{fig:Req1} 
Spatiotemporal patterns for $R=1$. Spatial 2D $X-$profiles for: 
a) $\varphi =73^{\circ}$, b) $\varphi =74^{\circ}$, c) $\varphi =120^{\circ}$, 
d) $\varphi =180^{\circ}$,
e) $\varphi =200^{\circ}$, f) $\varphi =310^{\circ}$, g) $\varphi =314^{\circ}$ and 
h) $\varphi =315^{\circ}$. The x-axis correspond to the horizontal arrangement
of the Brusselators and the y-axis to their vertical arrangement on the
2D square lattice network.
 Other parameters as in Fig.~\ref{fig:02}.
All simulations start from the same
random initial conditions.
}
\end{figure}

 Of special interest are the patterns developed for $R=1$ a
nd with various $\varphi$ values. 
Note that the case $R=1$ does not strictly correspond to the 
discretized regular diffusion, due to 
the influence of the rotational matrix $\varphi$. 
Only the case $\varphi =0, \>\> R=1$
can be considered as regular diffusion. The results for incremental values of $\varphi$
are shown in Fig.~\ref{fig:Req1}a-h.

\par For small $\varphi \le 73^{\circ}$ and large  $\varphi \ge 314^{\circ}$ values
of the rotation angle, the system presents coherence as shown in 
Fig.~\ref{fig:Req1}a and h. Between these values a certain degree of incoherence
reins in the system as presented in panels b to g of the same figure.
The borderline values written here were obtained for
 the working parameter set given in the first
paragraph of section~\ref{sec:chimeras}. For different generic parameter values
the borders between full coherence and partial coherence (chimera states)
 are appropriately adjusted.

\par
\begin{figure}[h]
\includegraphics[width=0.40\textwidth,angle=0.0] {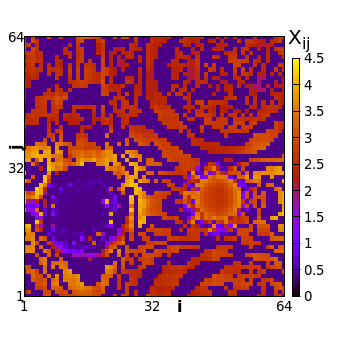}
\caption{\label{fig:two-coh-spots} 
Spatial 2D $X-$profiles for the two asymmetric stable coherent spots occurring at : 
 $(\varphi =91^{\circ}, \>\> R=10)$.
 Other parameters as in Fig.~\ref{fig:02}.
All simulations start from the same
random initial conditions.
}
\end{figure}

 Finally, 
a notable stable pattern was recorded in the vicinity of the parameter regions
where the coherent stripes were found, for $87^{\circ}  \le \varphi \le 92^{\circ}$ 
and
$5 \le R \le 12$. An example is provided in Fig.~\ref{fig:two-coh-spots} 
for ($\varphi =91^{\circ},\> R=10$); 
it consists of two asymmetric coherent spots.
The asymmetric spots oscillate out of phase, as one can see from the present snapshot
where at the same time the $X$-variable of the largest spot is $\approx 0.5$ (blue
color) and $X\approx 3$ (orange color) for the smaller one. Traces of similar patterns
are also seen in the surroundings of this parameter value.
  
%

\end{document}